\documentclass[superscriptaddress,nofootinbib,amsmath,amssymb,aps,twocolumn]{revtex4-2}

\usepackage{graphicx,enumerate,xcolor,enumitem,xspace,braket}
\usepackage{ulem}
\usepackage[colorlinks,citecolor=blue]{hyperref}
\usepackage{amsthm}
\usepackage{amsmath}
\usepackage{amssymb}
\usepackage{mathrsfs}

\newcommand{\nn}{\nonumber\\}

\newcommand{\av}[1]{\left\langle #1 \right\rangle}
\newcommand{\hatSij}[1]{\ensuremath{\hat{S}_{\{#1\}}}}
\newcommand{\Sij}[1]{\ensuremath{S_{\{#1\}}}}
\newcommand{\Iobs}[1]{\ensuremath{\mathcal{I}_3^{(#1)}}}
\newcommand{\Proj}[1]{\ensuremath{\hat{\Pi}_{\ket{#1}}}}
\newcommand{\tr}{\operatorname{Tr}}
\newcommand{\gev}{\operatorname{GeV}}

\newcommand{\fb}{\operatorname{fb}}
\newcommand{\dO}{\operatorname{d\Omega}}
\newcommand{\dtheta}{\operatorname{d\theta}}
\newcommand{\dphi}{\operatorname{d\varphi}}
\newcommand{\madgraph}{\textsc{MadGraph5\_aMC@NLO}~\cite{madgraph}\xspace}
\newcommand{\pythia}{\textsc{Pythia8}~\cite{pythia}\xspace}
\newcommand{\fastjet}{\textsc{Fastjet}~\cite{fastjet}\xspace}

\newcommand{\frakq}{\mathfrak{q}}
\newcommand{\frakn}{\mathfrak{n}}

\newcommand{\newobs}{{linear polarization}\xspace}
\allowdisplaybreaks[4]

\begin{document}

\title{New observables for testing Bell inequalities in $W$ boson pair production}

\author{Qi Bi}
\email{biqi@ihep.ac.cn}
\affiliation{Theoretical Physics Division, Institute of High Energy Physics, Chinese Academy of Sciences, Beijing 100049, China}
\affiliation{School of Physics, University of Chinese Academy of Sciences, Beijing 100049, China}

\author{Qing-Hong Cao}
\email{qinghongcao@pku.edu.cn}
\affiliation{School of Physics, Peking University, Beijing 100871, China}
\affiliation{Center for High Energy Physics, Peking University, Beijing 100871, China}

\author{Kun Cheng}
\email{chengkun@pku.edu.cn}
\affiliation{School of Physics, Peking University, Beijing 100871, China}

\author{Hao Zhang}
\email{zhanghao@ihep.ac.cn}
\affiliation{Theoretical Physics Division, Institute of High Energy Physics, Chinese Academy of Sciences, Beijing 100049, China}
\affiliation{School of Physics, University of Chinese Academy of Sciences, Beijing 100049, China}
\affiliation{Center for High Energy Physics, Peking University, Beijing 100871, China}

\begin{abstract}
 We show that testing Bell inequalities in $W^\pm$ pair systems by measuring their angular correlation suffers from the ambiguity in kinetical reconstruction of the di-lepton decay mode. We further propose a new set of Bell observables based on the measurement of the linear polarization of the  $W$ bosons that can be used in the semi-leptonic decay mode of $W^\pm$ pair, and we analyze the prospects of testing the violation of Bell inequalities at $e^+ e^-$ colliders.
\end{abstract}

\maketitle
\section{Introduction}
Quantum entanglement is a characteristic property of quantum states, and many criteria to determine the quantum entanglement have been developed, such as Bell inequalities~\cite{Bell:1964kc,Clauser:1969ny}, partial transposition~\cite{Peres:1996dw,Horodecki:1997vt} and concurrence~\cite{Hill:1997pfa,Wootters:1997id}. Among those criteria, the Bell inequality is based on directly measuring the non-locality of observables, and therefore of more experimental concern. Low energy experimental tests of the Bell inequalities have been performed in many quantum systems, such as photon pair~\cite{PhysRevLett.28.938,PhysRevLett.49.1804,PhysRevLett.89.240401} and superconducting systems~\cite{ansmann2009violation,zhong2019violating}, and many achievements have been made to avoid possible loopholes when testing local realism in these experiments~\cite{Hensen:2015ccp,giustina2015significant,shalm2015strong,storz2023loophole}.
In recent years, the test of entanglement at high-energy colliders draws more attention. 
It is proposed to test entanglement in many bi-particle systems produced at colliders, including the $t\bar{t}$ pair~\cite{Afik:2020onf,Fabbrichesi:2021npl,Severi:2021cnj,Aguilar-Saavedra:2022uye,Aoude:2022imd}, $W^+W^-$ pair~\cite{Barr:2021zcp,Aguilar-Saavedra:2022mpg,Ashby-Pickering:2022umy,Fabbrichesi:2023cev,Aoude:2023hxv}, $ZZ$ pair~\cite{Aoude:2023hxv,Ashby-Pickering:2022umy,Fabbrichesi:2023cev,Aguilar-Saavedra:2022wam} and $tW$ system \cite{Aguilar-Saavedra:2023hss}.

While the violation of Bell inequalities has already been confirmed in qubit systems, tests of Bell inequalities in massive vector boson pair systems, the only fundamental qutrit systems in our nature, are still pending. It is fascinating to check the violation of Bell inequalities in entangled ``quNit'' systems with $N\geqslant 3$ since the results for large $N$ are shown to be more resistant to noise with a suitable choice of the observables \cite{PhysRevLett.85.4418,PhysRevA.64.024101,PhysRevA.64.052109,CGLMP}.  Although the $W^\pm$ pair system is shown to be theoretically more promising than $Z$ pair to test the 3-dimensional Bell inequalities~\cite{Ashby-Pickering:2022umy,Fabbrichesi:2023cev}, a feasible experimental approach to test the Bell inequalities in $W^\pm$ system is yet to be proposed. In previous studies~\cite{Barr:2021zcp,Aguilar-Saavedra:2022mpg,Ashby-Pickering:2022umy,Fabbrichesi:2023cev}, it was a  common practice to use the di-lepton decay mode of $W^\pm$ pair to test the entanglement, because the complete density matrix $\hat\rho_{WW}$ of $W^\pm$ system can be reconstructed in this channel then all entanglement criteria can be calculated from $\hat\rho_{WW}$ directly. But it was also pointed out that a two-fold ambiguity in the kinetic reconstruction is unavoidable in the di-lepton decay mode of $W^\pm$. In this work, we show that this ambiguity in the di-lepton decay mode can lead to a fake signal of the violation of Bell inequalities. Therefore, it is necessary to search for an alternative approach to test the Bell inequalities in the $W^\pm$ system.

For $W^\pm$ pairs produced at electron-positron colliders, an event-by-event kinetical reconstruction of $W^\pm$ pair can be performed without any ambiguity in their semi-leptonic decay mode. Though the semi-leptonic decay mode is a major channel to measure $W^\pm$ spin correlations at the LEP~\cite{Gounaris:1991ce,Gounaris:1992kp,OPAL:2000wbs,L3:2003tnr}, this decay mode was not used to test entanglement before due to the loss of angular momentum information of the $W$ boson as it is hard to identify the flavor of light jets. Therefore, the conventional approach to test Bell inequalities, which relies on measuring angular momentum correlations between $W^+$ and $W^-$, cannot be applied in the semi-leptonic decay mode.  Although only partial information on the density matrix $\hat\rho_{WW}$ can be reconstructed in the semi-leptonic decay mode of the $W^\pm$ pairs, we succeed in finding a new observable to test the Bell inequalities. More specifically, we construct new Bell observables based on the \newobs of $W$ bosons, which does not require tagging the flavor of the decay products of $W^\pm$ pairs.
We show that these new Bell observables can be correctly measured from the semi-leptonic decay of $W^\pm$ pairs, providing a feasible way to test Bell inequalities in $W^\pm$ pair production.

\section{Method}

We begin by introducing the theoretical framework of testing Bell inequalities in $W^\pm$ pairs. Ignoring the interactions between the $W^\pm$ bosons, the $W^\pm$ pair system can be described by the tensor product Hilbert space $\mathscr H=\mathscr H_A\otimes\mathscr H_B$ of the state Hilbert space $\mathscr H_A$ of $W^+$ and the state Hilbert space $\mathscr H_B$ of $W^-$. 
Fixing the momentum of the $W^\pm$ boson, the subspace $\mathcal H_{A/B}$ is 3-dimensional representation space of the rotation group $SU(2)$.
Considering some measurements $\hat{A}_i$ and $\hat{B}_i$ carried out in these 3-dimensional spin spaces of $\mathcal{H}_A$ and $\mathcal{H}_B$, their outcomes $A_{i}$ and $B_{i}$ have three possible values in $\{-1,0,1\}$, where the index $i=1,2$ is used to denote different measurements on the same system.
The optimal~\cite{Masanes:2002} generalized Bell inequality for 3-dimensional systems, also referred as Collins-Gisin-Linden-Massar-Popescu (CGLMP) inequality~\cite{CGLMP}, states that the upper limit of the following expression,
\begin{align}\label{eq:I3}
    \mathcal{I}_3\equiv &+ \big[P\left(A_1=B_1\right)+P\left(B_1=A_2+1\right) \nn
    &\quad+P\left(A_2=B_2\right)+P\left(B_2=A_1\right)\big] \nn
    & -\big[P\left(A_1=B_1-1\right)+P\left(B_1=A_2\right) \nn
    & \quad+P\left(A_2=B_2-1\right)+P\left(B_2=A_1-1\right)\big],
\end{align}
is 2 for any local theory, i.e., $\mathcal{I}_3\leq 2$. Here, $P(A_i=B_j+k)$ denotes the probability that the measurement outcomes $A_i$ and $B_j$ differ by $k$ modulo 3.

For a non-local theory, the inequality $\mathcal{I}_3\leq 2$ no longer holds, and the upper limit of $\mathcal{I}_3$ is 4 instead. In other words, as long as there exists a set of measurements such that the corresponding CGLMP inequality is violated, i.e.,
\begin{align}\label{eq:maxI3AABB}
    \max_{\hat{A}_1,\hat{A}_2,\hat{B}_1,\hat{B}_2} \mathcal{I}_3(\hat{A}_1,\hat{A}_2;\hat{B}_1,\hat{B}_2)> 2,
\end{align}
the non-locality of the system is confirmed.

A direct way to evaluate $\mathcal{I}_3$ is to project the density matrix $\hat\rho_{WW}$ to the eigenstates of the operators $\hat{A}_i$ and $\hat{B}_i$, e.g., the first term in Eq.~\eqref{eq:I3} is
\begin{align}
    P(A_1=B_1)= \sum_{\lambda=-1}^1 & \tr\left[ \hat\rho_{WW} \Proj{A_1=\lambda,B_1=\lambda } \right],
\end{align}
where $\Proj{\psi}\equiv\ket{\psi}\bra{\psi}$ is the projection operator. At lepton colliders, $\hat\rho_{WW}$ could be theoretically calculated with the transition amplitudes $\mathcal M_{WW}$ of the $e^+e^-\to W^+W^-$ process in the electroweak standard model (SM) to 
\begin{equation}
\hat\rho_{WW}\propto\mathcal M_{WW} \hat\rho_{ee}\mathcal M_{WW}^\dagger,
\end{equation}
where $\mathcal M_{WW}$ is a $9\times 4$ matrix in spin space, and $\hat\rho_{ee}$ is the $4\times 4$ spin density matrix of the initial state $e^+e^-$ which is $\hat{I}_4/4$ for unpolarized beam. Here, $\hat I_d$ is the identity operator in $d$-dimensional Hilbert space. Unfortunately, the spin state of the $W$ bosons could not be directly measured at colliders. Therefore, we next introduce how to obtain $\rho_{WW}$ from the decay products of $W^\pm$ pairs.

As a preliminary, we start with the spin density matrix of one $W$ boson, which could be generally parameterized as
\begin{equation}
    \label{eq:rhoWSij}
    \hat\rho_W=\frac{\hat I_3}{3}+\sum_{i=1}^3 d_i \hat S_i + \sum_{i,j=1}^3 q_{ij} \hat S_{\{ij\}},
\end{equation}
where $\hat S_i$ is the $i$-th component of the 3-dimensional angular momentum operator, $\hat S_{\{ij\}}\equiv \hat S_i\hat S_j+\hat S_j \hat S_i$, and the coefficients $q_{ij}$ is symmetric traceless.
Note that the two sets of operators $S_i$ and $\Sij{ij}$ are orthogonal, i.e., $\tr(S_i S_{\{jk\}})=0$.\footnote{For more properties of this parametrization and the relations of the operators $\hat S_i$'s and $\hat S_{\{ij\}}$'s, see Appendix~\ref{app:SiSij} for details.}
The parametrization separates the information of angular momentum and linear polarization of the $W$-boson explicitly. 
On the one hand, the expectation value of the angular momentum of the $W$-boson along direction $\vec{a}$ yields $\tr(\hat{\vec{S}}\cdot\vec{a}~\hat\rho_W)=2\vec{d}\cdot\vec{a}$, which only depends on $d_i$.  On the other hand, a (partly) linear polarized $W$-boson has zero angular momentum with $d_i=0$, and its polarization information only depends on $q_{ij}$.

With the polarization information of each term of $\hat\rho_{W}$ in mind, we continue to reconstruct the density matrix of a $W$ boson from its decay products.
In its rest frame, ignoring the tiny mass of the final state fermion and anti-fermion, the $W$ boson always decays into a negative helicity fermion and a positive helicity anti-fermion since the weak interaction only couples to left-handed fermions, and we denote the normalized direction of outgoing anti-fermion in the rest frame of the $W$ boson as $\vec{\frakn}$, which is just the direction of the (experimentally {\it{measured}}) total angular momentum. In additional to $\frakn_i$, we define a symmetric and traceless tensor of rank-2 (the quadrupole)
\begin{equation}\label{eq:qijdefinition}
    \frakq_{ij}\equiv \frakn_i\frakn_j - \frac{1}{3}\delta_{ij}
\end{equation}
to describe the high-order information on the distribution of decay products. The probability of finding an anti-fermion in an infinitesimal solid angle $\dO$ of direction $\vec{\frakn}(\theta,\phi)$ from the $W$-boson decay products is~\cite{Barr:2021zcp}
\begin{equation}\label{eq:pnrho}
    p(\vec{\frakn};\hat\rho_W)=\frac{3}{4\pi}\tr\left[\hat\rho_W \hat{\Pi}_{\vec{\frakn}}\right],
\end{equation}
where the projection operator $\hat{\Pi}_{\vec{\frakn}}$ selects the positive helicity anti-fermion in the direction $\vec{\frakn}$. The explicit expression of $p(\vec{\frakn};\rho_W)$ is shown in Appendix~\ref{app:densitymatrix}.

By integrating the probability with the kinetic observables $\frakn_i$ and $\frakq_{ij}$, it is found that the parameters $d_i$ and $q_{ij}$ in Eq.~\eqref{eq:rhoWSij} are directly determined by the averages of these kinetic observables,
\begin{align}\label{eq:average1}
    d_i = \langle\frakn_i\rangle,\quad
    q_{ij} = \frac{5}{2}\langle\frakq_{ij}\rangle,
\end{align}
which are defined as
\begin{align}
    \label{eq:intgrateni}
    \langle\frakn_i\rangle&\equiv \int \frakn_i ~p(\vec{\frakn};\hat\rho_W) \dO,\\
    \label{eq:intgrateqij}
    \langle\frakq_{ij}\rangle&\equiv \int \frakq_{ij} ~p(\vec{\frakn};\hat\rho_W) \dO.
\end{align}
Therefore, the parameters $d_i$'s, which are related to the angular momentum of the $W$ boson, are determined by  $\langle\frakn_i\rangle$, the dipole distributions of the anti-fermion, and require distinguishing fermion from anti-fermion (or flavor tagging).
The parameters $q_{ij}$'s, which are related to the \newobs of the $W$ boson, are determined by $\langle\frakq_{ij}\rangle$, the quadrupole distributions of the decay products, and do not need flavor tagging.

Likewise, the density matrix of $W^\pm$ pair can be reconstructed from the distribution of their decay products. The density matrix $\hat\rho_{WW}$ is parameterized as (see Ref.~\cite{Rahaman:2021fcz} for a similar formulation)
\begin{align}\label{eq:rhoWWSij}
    \hat\rho_{WW}=&\frac{\hat I_9}{9} + \frac{1}{3} d_i^+ \hat S_i^+\otimes \hat I_3 +\frac{1}{3} q_{ij}^+ \hat S_{\{ij\}}^+\otimes \hat I_3 \nn
     &+\frac{1}{3} d_i^- \hat I_3\otimes \hat S_i^- +\frac{1}{3} q_{ij}^- \hat I_3\otimes \hat S_{\{ij\}}^-  \nn
     &+ C^d_{ij}\hat S_i^+\otimes \hat S_j^- + C^q_{ij,k\ell}\hat S_{\{ij\}}^+\otimes \hat S_{\{k\ell\}}^-\nn
     &+ C^{dq}_{i,jk}\hat S_i^+\otimes \hat S_{\{jk\}}^- + C^{qd}_{ij,k} \hat S_{\{ij\}}^+\otimes \hat S_k^-,
\end{align}
where $\hat S_i^+$ ($\hat S_i^-$) and $\hat S^+_{\{ij\}}$ ($\hat S^-_{\{ij\}}$) is the  $\hat S_i$ and $\hat S_{\{ij\}}$ operator defined in the rest frame of the $W^+$ ($W^-$) boson, respectively, and the repeated indices are summed as in Eq.~\eqref{eq:rhoWSij}. 

We use $\vec{\frakn}^\pm$ to denote the normalized directions of two outgoing anti-fermions decayed from $W^\pm$ in the rest frame of $W^\pm$, respectively. The quadrupole kinetic observables $\frakq_{ij}^\pm\equiv \frakn_i^\pm\frakn_j^\pm - \frac{1}{3}\delta_{ij}$ are defined similarly.
Again, all the parameters in $\hat\rho_{WW}$ can be reconstructed from the average of the observables $\frakn_i^\pm$, $\frakq_{ij}^\pm$ and their correlations. With a detailed calculation in Appendix~\ref{app:densitymatrix}, we enumerate the kinetic observables needed to obtain each term of $\hat \rho_{WW}$ as follows:

The first two lines of Eq.~\eqref{eq:rhoWWSij} are determined by the decay products distributions of each $W$ boson itself,
\begin{align}\label{eq:average2_d}
    d^\pm_{i}&=\av{\frakn^\pm_i}, \\
    q^\pm_{ij}&=\frac{5}{2}\av{\frakq^\pm_{ij}}.
\end{align}
The terms in the third line of Eq.~\eqref{eq:rhoWWSij} are determined by the correlations between the dipole or quadrupole distributions of the decay products of $W^+$ and $W^-$.
\begin{align}\label{eq:average2_dd}
    C^d_{ij}&=\av{\frakn^+_i \frakn^-_j}, \\
    C^q_{ij,k\ell}&=\frac{25}{4}\av{\frakq^+_{ij} \frakq^-_{k\ell}}.
\end{align}
The terms in the fourth line of Eq.~\eqref{eq:rhoWWSij} are determined by the correlations between the dipole distribution of the decay products of one $W$ boson and the quadrupole distribution of the decay products of the other.
\begin{align} \label{eq:average2_dq}
    C^{dq}_{i,jk}&=\frac{5}{2}\av{\frakn^+_{i}\frakq^-_{jk}},\\
    \label{eq:average2_qd}
    C^{qd}_{ij,k}&=\frac{5}{2}\av{\frakq^+_{ij}\frakn^-_{k}}.
\end{align}

With Eqs.~\eqref{eq:average2_d}-\eqref{eq:average2_qd}, we are ready to obtain the complete density matrix $\hat\rho_{WW}$ of the system and test the Bell inequalities. Besides, it is worth emphasizing that tagging the flavor of the decay product $W^+$ or $W^-$ is necessary to fix the overall sign of $\frakn_i^+$ or $\frakn_i^-$, but not necessary to obtain $\frakq_{ij}^\pm$.

\section{Neutrino reconstruction in di-lepton decay mode}

As a usual practice, the Bell inequalities in $W^\pm$ system are tested by measuring the angular momentum correlations of the two $W$ bosons. In that case, the operators in Eq.~\eqref{eq:maxI3AABB} are chosen as angular momentum operators and the Bell observable $\mathcal{I}_3$ is defined as
\begin{align}
    \Iobs{S}\equiv\mathcal{I}_3(\hat{S}_{\vec{a}_1},\hat{S}_{\vec{a}_2};\hat{S}_{\vec{b}_1},\hat{S}_{\vec{b}_2} ),
\end{align}
where $\hat{S}_{\vec{n}}\equiv \hat{\vec{S}}\cdot \vec{n}$, and $(\vec{a}_i,\vec{b}_i)$ are a set of directions in the rest frames of $W^\pm$ respectively, and the maximum of $\Iobs{S}$ is obtained by scanning all possible directions $\vec{a}_i$ and $\vec{b}_i$ to measure the angular momentum.

To measure the angular momentum of each $W$-boson, the $\hat S_i$ dependent terms of $\hat\rho_{WW}$ such as $C^d_{ij}\hat S_i\otimes \hat S_j$ must be correctly obtained. Since these terms are reconstructed from the kinetic observable $\frakn^\pm_i$, distinguishing fermion from anti-fermion in both $W$ boson decay processes is necessary.
In the hadronic decay mode of $W$ boson, it is shown that the jet substructures such as jet charge can help to distinguish light quark flavor, but the tagging efficiency is still very low~\cite{Li:2023tcr}. Therefore, only di-lepton decay mode, $W^+(\to\ell^+\nu_\ell)W^-(\to\ell^-\bar{\nu}_\ell)$, is considered in previous studies to calculate the criteria of entanglement~\cite{Barr:2021zcp,Aguilar-Saavedra:2022mpg,Ashby-Pickering:2022umy,Fabbrichesi:2023cev}. 

However, in the di-lepton decay mode of $W^\pm$, there are two undetectable neutrinos and the momenta of $W^\pm$ cannot be obtained directly. To reconstruct the rest frame of $W^\pm$ and obtain $\frakn_{i}^{\pm}$ and $\frakq_{ij}^{\pm}$, the neutrino momenta must be solved from two observed leptons using on-shell conditions, but the solution suffers from twofold discrete ambiguity~\cite{Hagiwara:1986vm} even if we ignore $W$ boson width and experimental uncertainties. In other words, the false solutions behave like irreducible backgrounds that are comparable with signals. As a result,  attempting to measure the theoretical value of \Iobs{S} calculated in previous studies are subject to experimental difficulties in kinetical reconstruction.

\begin{figure}
    \includegraphics[width=.95\linewidth]{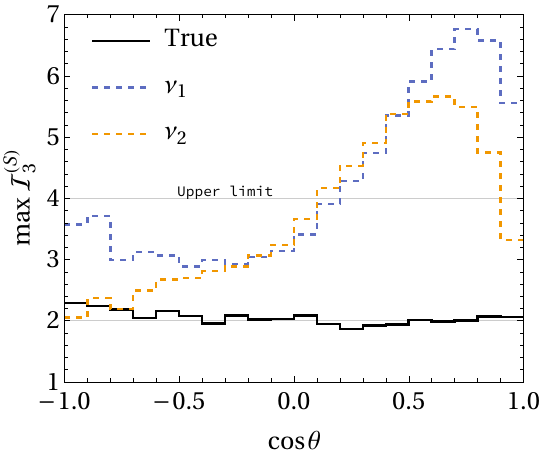}
    \caption{The maximum value of \Iobs{S} calculated with true neutrino momentum (solid line) or solved neutrino momentum (dashed lines) at $\sqrt s=240 \gev$ electron-positron collider. Here, $\theta$ is the scattering angle between $W^+$ and incoming $e^+$ beam, $\nu_1$ or $\nu_2$ denotes the neutrino solution with larger or smaller transverse momentum respectively.}
    \label{fig:WWlvlv}
\end{figure}
To illustrate the impact of the twofold ambiguity, we use the unpolarized scattering process $e^+e^-\to W^+ W^-$ with $\sqrt{s}=240\gev$ as an example. We perform a parton level simulation using \madgraph with full spin correlations and Breit-Wigner effects included.  
From the momenta of the detected leptons, we obtain two degenerate solutions of the neutrino momentum that satisfying the kinetic conditions~[39].  We choose the solution with a larger or smaller transverse momentum ($\nu_1$ and $\nu_2$ in Fig.~\ref{fig:WWlvlv}, respectively) to reconstruct the rest frame of $W^\pm$ and then reconstruct the density matrix from Eqs.~\eqref{eq:average2_d}-\eqref{eq:average2_qd}. When averaging the kinetic observables in Eqs.~\eqref{eq:average2_d}-\eqref{eq:average2_qd}, we choose to work in the beam basis~\cite{Mahlon:1995zn,Afik:2020onf}, where $\hat{z}$ is along the incoming $e^+$ beam direction, $\hat{x}\propto \hat{z}\times \vec{p}_{W^+}$ is the normal direction of the scattering plan, and $\hat{y}=\hat{z}\times\hat{x}$. For comparison, we also include the results calculated with the knowledge of the true momentum of each neutrino, as shown in Fig.~\ref{fig:WWlvlv}.
For a better illustration, the results in Fig.~\ref{fig:WWlvlv} is reconstructed from the parton-level momenta of leptons directly without any selection cuts, so that the two fold ambiguity makes the only difference between the reconstructed results and the true results. It is found that the twofold ambiguity is destructive for testing Bell inequalities with $\Iobs{S}$, as the observed value of $\Iobs{S}$ can be much larger than its theoretical value and may even exceed the physical upper limit, indicating a fake signal of entanglement. Considering momentum smearing effect and kinetic cuts further obscure the test of Bell inequalities.

Therefore, it is shown that the experimentally observed \Iobs{S} cannot directly represent the entanglements between the $W^\pm$ pair. In addition, other entanglement criteria that can only be measured at full leptonic decay channel of $W^\pm$ pair, such as the concurrence and partial trace, also suffer from the two-fold solutions of neutrino momentum.
The ambiguity of neutrinos also exists in the more studied $t\bar t$ case, where some reconstruction techniques such as unfolding~\cite{CMS:2019nrx,Severi:2021cnj,Dong:2023xiw} and parameter fitting~\cite{ATLAS:2023,Han:2023fci} are commonly used.\footnote{The parameter fitting is argued to be more trusty than unfolding, see, e.g., Ref.~\cite{Kuusela2012,Han:2023fci}.}
Similarly, to test Bell inequality in $W^\pm$-pair system using \Iobs{S}, some reconstruction techniques are also necessary.  In this work, instead of digging into the technique details, we find that the we can simplify the test of Bell inequality in $W^\pm$ pair with a new observable in the semi-leptonic decay mode.

\section{New observables in semi-leptonic decay mode}

\begin{figure}
    \includegraphics[width=.49\linewidth]{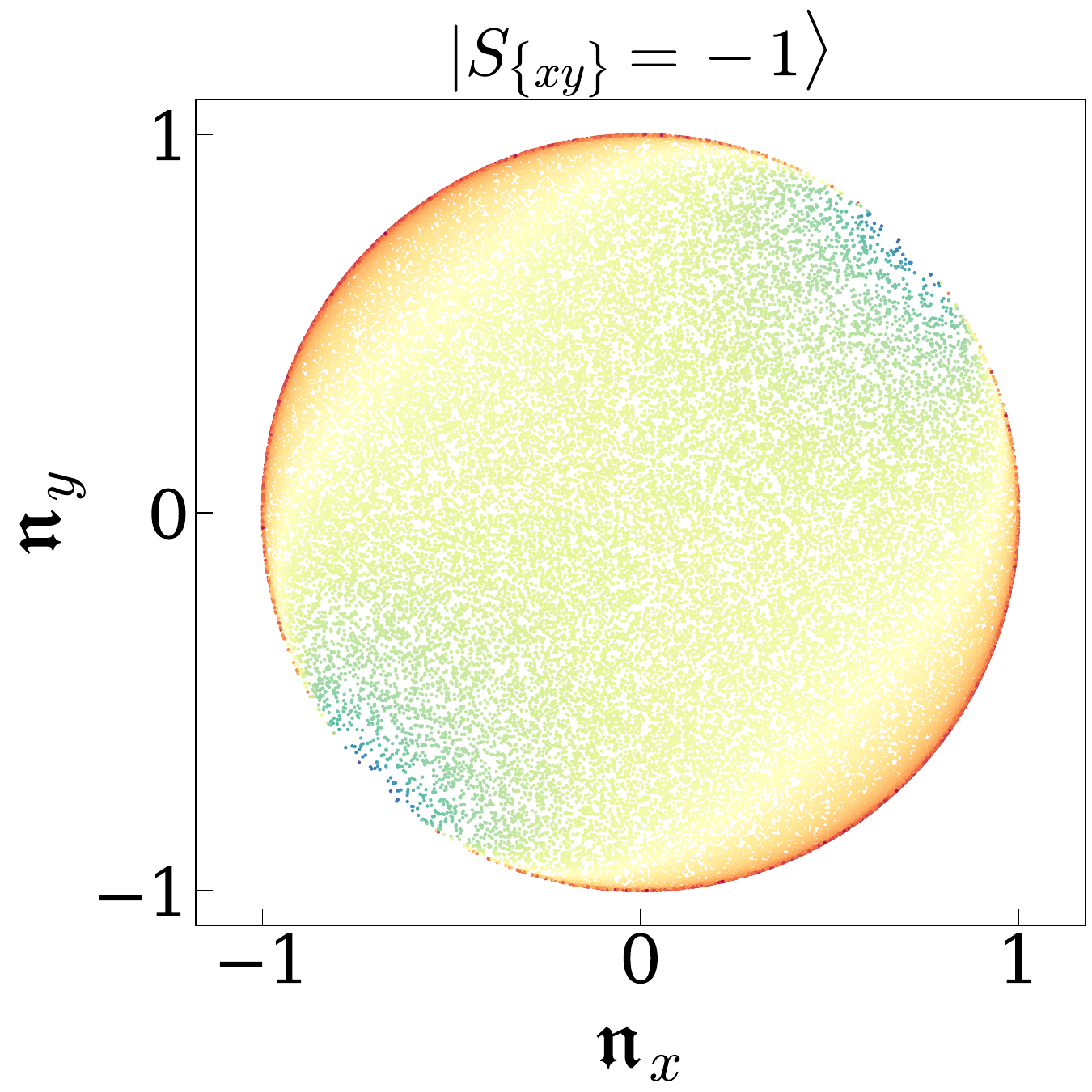}
    \includegraphics[width=.49\linewidth]{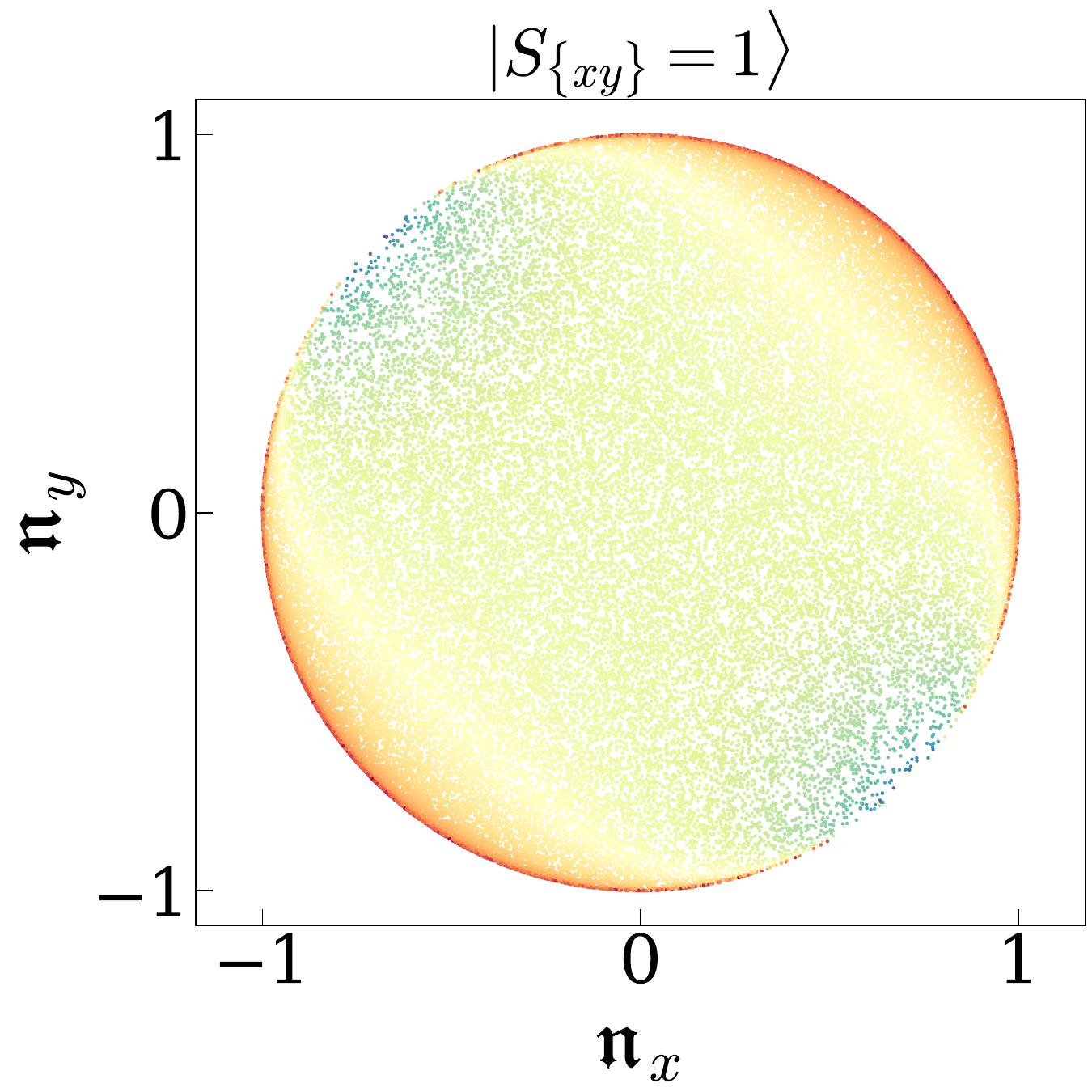}
    \caption{Distributions of the decay products of $W$ bosons in different eigenstates of \Sij{xy}, viewed from the $z$-direction. The color stands for the density of distribution. The decay products of the $W$ boson in the state $\ket{\Sij{xy}=\pm1}$ have positive or negative quadrupole distribution respectively.}
    \label{fig:eigSxy}
\end{figure}

In the semi-leptonic decay modes of $W^\pm$ pair produced at lepton colliders, all momenta can be determined without any ambiguity. Despite the convenience in kinetical reconstruction in the semi-leptonic decay modes, a complete density matrix $\rho_{WW}$ cannot be reconstructed in these modes, because the angular momentum of the $W$-boson decaying to hadrons cannot be measured without jet flavor tagging. Consequently, the Bell observable \Iobs{S} is not valid in these decay channels. However, the \newobs of the $W$-boson decaying to hadrons can still be measured correctly, because the \newobs of a $W$-boson is determined from the quadrupole distribution $\av{\frakq_{ij}}$ of its decay products, which does not depend on the overall sign of $\vec\frakn$. To construct a Bell observable that can be measured in the semi-leptonic decay mode of $W^\pm$, we choose operator $\hatSij{xy}\equiv \{\hat{S}_x,\hat{S}_y\}$ to measure the \newobs of the $W$-boson decaying to hadrons. Note that the eigenstates $\ket{\Sij{xy}=\pm 1}$ are purely linear polarized states with different polarization directions on the $xy$-plane,
\begin{align}\label{eq:eigSxy}
    \vec{\epsilon}_{\ket{\Sij{xy}=-1}}&=\frac{1}{\sqrt{2}}(1,1,0),\nn
    \vec{\epsilon}_{\ket{\Sij{xy}=1}}&=\frac{1}{\sqrt{2}}(1,-1,0),\nn
    \vec{\epsilon}_{\ket{\Sij{xy}=0}}&=(0,0,1),
\end{align}
and the expectation value of $\hatSij{xy}$, $E(\hatSij{xy})$, is directly determined by the quadrupole distribution of the decay products with $E(\hatSij{xy})=10\av{\frakq_{xy}}$, as shown in Fig.~\ref{fig:eigSxy}.

We first consider the decay channel $W^+(\to \ell^+ \nu_\ell)W^-(\to jj)$, where the lepton $\ell$ is electron or muon. In this channel, both the angular momentum of $W^+$ and the linear polarization of $W^-$ can be determined correctly. Therefore, we choose to measure the correlation between the angular momentum of $W^+$ and the \newobs of $W^-$ to test the Bell inequalities in this channel, and the new Bell observable is defined as
\begin{equation}\label{eq:ISL}
    \Iobs{S,L}\equiv \mathcal{I}_3(\hat{S}_{\vec{a}_1},\hat{S}_{\vec{a}_2};\hatSij{x_3 y_3},\hatSij{x_4 y_4}), 
\end{equation}
where $(x_i,y_i)$ are the coordinates in the rest frame of $W^-$, and $\vec{a}_i$ are the directions in the rest frame of $W^+$.

The observable $\Iobs{S,L}$ is reconstructed in following steps.  
First, measure the distribution of $W^\pm$ decay products and obtain the parameters of the density matrix using Eqs.~(12)-(17).  The $W^-$ decays hadronically and only the quadruple distribution of it is needed.
Second, construct the density matrix $\hat\rho_{WW}$ in Eq.~(11) with the parameters obtained in the previous step.
Third, calculate the probabilities $P(S_{\vec a}=\Sij{xy})$ and $P(S_{\vec a}=\Sij{xy}\pm 1)$ by projecting the density matrix to the eigenstates
$\ket{S_{\vec{a}}=A}\otimes\ket{\Sij{xy}=B},(A,B=-1,0,1)$
\footnote{See Appendix~\ref{app:observable} for examples of explicit expression of the projection results.}, and then construct \Iobs{S,L} according to Eq.~(1).
Note that the coefficients related to the angular momentum of $W^-$, namely $d^-_i$, $C^d_{ij}$ and $C^{qd}_{ij,k}$ are independent of the projection, which is the reason why they are not needed in the first step.

\begin{figure}
    \includegraphics[width=.95\linewidth]{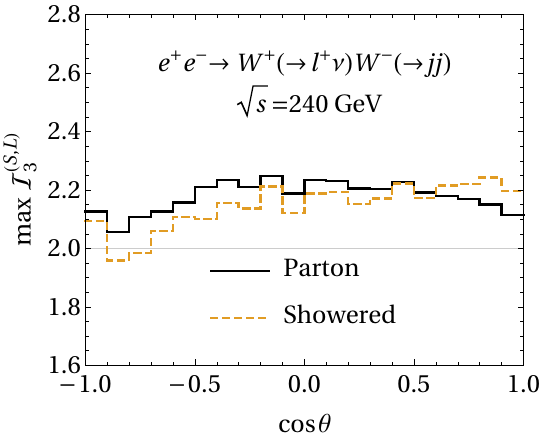}
    \caption{The value of \Iobs{L,S} for $W^\pm$ pair produced from $e^+ e^-\to W^+ W^-$ with $\sqrt{s}=240\gev$.}
    \label{fig:WWjjlv}
\end{figure}

We perform a Monte-Carlo simulation of $e^+e^- \to W^+(\to \ell^+ \nu_\ell)W^-(\to jj)$ processes with $\sqrt{s}=240\gev$.  The parton level events are generated by \madgraph and then passed to \pythia for showering and hadronization. The showered events are passed to \fastjet for jet clustering with Durham algorithm, and the clustering is taken to be stopped when it reaches 2 exclusive jets.
We require the energy of lepton and jets to be larger than $15\gev$ and $5\gev$ respectively, and the invariant mass of the two jets satisfy $|m_{jj}-m_W|<20\gev$.
In addition, we require the angle between the lepton missing vector, $\theta_{\ell p_{\rm miss}}$, to satisfy $\cos\theta_{\ell p_{\rm miss}}<0.2$~\cite{OPAL:1997qhh}, so that the background from $W\to \tau\nu$ are negligible.
As shown in Fig.~\ref{fig:WWjjlv}, we find that the showering and selection cuts slightly dilute the signal of entanglements, but the observed \Iobs{S,L} is still in good consistency with the parton level results, making \Iobs{S,L} a good observable to test Bell inequalities in $W^\pm$ pair system. The statistical significance of observing the violation of the Bell inequalities can be calculated with the standard $\chi^2$ statistical test,
\begin{equation}
    \chi^2 = \sum_i \left(\frac{\Iobs{S,L}-2}{\delta_i}\right)^2,
\end{equation}
where the sum runs over the bins with $\mathcal{I}_3>2$, and the statistical uncertainty $\delta_i$ are calculated from the \textit{standard error of mean} in Eqs.~\eqref{eq:average2_d}-\eqref{eq:average2_qd}. 
Using the observable \Iobs{S,L}, the Bell inequality violation can be tested at $3\sigma$ significance at $240\gev$ $e^+e^-$ collider with $150\fb^{-1}$ luminosity.

Likewise, another semi-leptonic decay mode, $W^+(\to jj)W^-(\to\ell^-\bar{\nu}_\ell)$, can also be used to test the Bell inequalities. In this decay mode, we choose to measure the \newobs of the $W^+$ and the angular momentum of $W^-$, and the Bell inequalities $\mathcal{I}_3\leq 2$ are tested by another observable,
\begin{align}
    \Iobs{L,S}\equiv \mathcal{I}_3(\hatSij{x_1 y_1},\hatSij{x_2 y_2};\hat{S}_{\vec{b}_1},\hat{S}_{\vec{b}_2}).
\end{align}
Combining the two semi-leptonic decay modes of $W^\pm$ pair produced at $240\gev$ $e^+e^-$ collider, one can verify the violation of the Bell inequality at $5\,\sigma$ significance with $180 \fb^{-1}$ integrated luminosity.

\section{Conclusion}

The commonly used criteria of entanglement rely on the di-lepton decay mode of $W^\pm$, because the di-lepton decay mode is the only decay mode that can be used to reconstruct the complete density matrix. However, we show that due to the irreducible ambiguity of neutrino momentum solutions in the di-lepton decay mode, testing entanglement in the di-lepton decay mode of $W^\pm$ pair may yield fake signals.

We provide a more realistic approach to test Bell inequalities in $W^\pm$ pair systems using a new set of Bell observables based on measuring the \newobs of $W$ bosons. Our observables depend on only part of the density matrix that can be correctly measured in the semi-leptonic decay mode of $W^\pm$. With these new Bell observables, it is found that the violation of Bell inequalities in $W^\pm$ pair produced at $240\gev$ electro-positron colliders can be tested at $5\sigma$ significance with an integrated luminosity of $180\fb^{-1}$. 

\begin{acknowledgments}    
We thank Yandong Liu, Tong Arthur Wu and Changlong Xu for their useful discussions. The work is supported in part by the National Science Foundation of China under Grants No. 11725520, No. 11675002, No. 12075257 and No. 12235001.
\end{acknowledgments}

\appendix

\section{Spin operators and their matrix representations}\label{app:SiSij}

In this appendix, we give some general properties of the spin operators and their matrix representation in the basis of the eigenstates of the angular momentum of the 3rd axis ($z$-axis).

The general spin operators $\hat S_i$ ($i=1,2,3$ or $x,y,z$) satisfy the angular commutation relation
\begin{equation}
    [\hat S_j,\hat S_k]=i\varepsilon_{jk\ell}\hat S_\ell,
    \label{eq:A1}
\end{equation}
where $\varepsilon_{jk\ell}$ is the 3-dimensional Levi-Civita symbol. In the 3-dimensional representation, the Casimir operator
\begin{equation}
    \hat S^2=\hat S_1^2+\hat S_2^2+\hat S_3^2=L(L+1)\hat I_{2L+1}=2\hat I_3,
    \label{eq:A2}
\end{equation}
where $L$ is the total angular momentum quantum number, and $L=1$ for vector boson spin operators.

With Eq (\ref{eq:A1}), we have
\begin{equation}
    \tr(\hat S_\ell)=-\frac{i}{2}\varepsilon_{jk\ell}\tr([\hat S_j,\hat S_k])=0,
\end{equation}
and
\begin{equation}
    \tr(\hat S_a\hat S_b)=-i\tr(\hat S_a\hat S_j\hat S_k)\varepsilon_{jkb}.
\end{equation}
When $a=b$, we have $\tr(\hat S_a^2)=-i\tr(\hat S_1\hat S_2\hat S_3)+i\tr(\hat S_1\hat S_3\hat S_2)=2$ for any $a$, so $\tr(\hat S_{\{11\}})=\tr(\hat S_{\{22\}})=\tr(\hat S_{\{33\}})=4\tr(\hat I_3)/3=4$. When $a\neq b$,
\begin{equation}
    \tr(\hat S_a\hat S_b)=-i\{\tr(\hat S_a\hat S_a\hat S_c)-\tr(\hat S_a\hat S_c\hat S_a)\}=0.
\end{equation}
So $\tr(\hat S_{\{12\}})=\tr(\hat S_{\{23\}})=\tr(\hat S_{\{31\}})=0$. With these results, we have
\begin{equation}
    \tr(\hat \rho_W)=1
\end{equation}
so that it is a normalized density matrix operator.

The vector representation of the angular momentum operator is special because $\hat I_3, \hat S_i, \hat S_{\{ij\}}$ gives a basis of the real linear space $\mathfrak A_3$ of all self-adjoint operators on 3-dimensional complex Hilbert space if we choose two linear combinations  $u_i\hat S_{\{ii\}}$ and $v_i\hat S_{\{ii\}}$ with  $\sum_iu_i=\sum_iv_i=0, u_i,v_i\in \mathbb{R}$. In fact, if we introduce a positive definite inner product $(\hat A,\hat B)\equiv\tr(\hat A^\dagger\hat B)$ in the operator space, they form an orthogonal basis. To verify this conclusion, we notice that
\begin{equation}
    \tr(\hat I_3^2)=3,
\end{equation}
so $\hat I_3/\sqrt 3$ is a normalized operator. For the angular operators, we have proved that $\tr(\hat S_i)=0$ and $\tr(\hat S_i\hat S_j)=2\delta_{ij}$. So $\{\hat I_3/\sqrt3,~ \hat S_i/\sqrt2\}$ is a set of orthonormal vectors. For $\hat S_{\{ij\}}$ ($i\neq j$), $\sum_iu_i\hat S_{\{ii\}}$ and $\sum_iv_i\hat S_{\{ii\}}$, we have proved that $\tr(\hat S_{\{ij\}})=4\delta_{ij}$, so with the constraints $\sum_iu_i=\sum_iv_i=0$ they are all orthogonal to $\hat I_3/\sqrt 3$. We first check whether they are orthogonal to $\hat S_i$. When $i\neq j\neq k$, without loss of generality, we assume that $(ijk)$ is an even permutation of $(1,2,3)$, then
\begin{eqnarray}
\tr(\hat S_i\hat S_{\{jk\}})&=&\tr(\hat S_i\hat S_j\hat S_k+\hat S_i\hat S_k\hat S_j)\nonumber\\
&=&-i\tr(\hat S_i\hat S_j\hat S_i\hat S_j-\hat S_i\hat S_j^2\hat S_i+\hat S_i^2\hat S_j^2\nonumber\\
&&-\hat S_i\hat S_j\hat S_i\hat S_j)\nonumber\\
&=&0.
\end{eqnarray}
When $j$ or $k$ is equal to $i$, without loss of generality, we assume $i=j\neq k$, then,
\begin{eqnarray}
\tr(\hat S_i\hat S_{\{ik\}})&=&\tr(\hat S_i\hat S_i\hat S_k+\hat S_i\hat S_k\hat S_i)\nonumber\\
&=&-i\varepsilon_{i\ell k}\tr(\hat S_i\hat S_i\hat S_i\hat S_\ell-\hat S_i\hat S_i\hat S_\ell\hat S_i+\hat S_i^2\hat S_\ell\hat S_i\nonumber\\
&&-\hat S_i\hat S_\ell\hat S_i\hat S_i)\nonumber\\
&=&0.
\end{eqnarray}
When $j=k$,
\begin{eqnarray}
\sum_jv_j\tr(\hat S_i\hat S_{\{jj\}})&=&2\sum_jv_j\tr(\hat S_i\hat S_j\hat S_j)\nonumber\\
&=&4v_i\tr(\hat S_i)+2\sum_{j\neq i}(v_j-v_i)\tr(\hat S_i\hat S_j^2)\nonumber\\
&=&2\sum_{j\neq i}(v_j-v_i)\tr(\hat S_i\hat S_j^2)\nonumber\\
&=&0.
\end{eqnarray}
Finally, we check the inner product between the $\hat S_{\{ij\}}$'s. Since there are only 3 possible values of the index, using the exchange symmetric property of the indices, we only need to check $\tr(\hat S_{\{ij\}}\hat S_{\{ik\}})$ and $\tr(\hat S_{\{ii\}}\hat S_{\{jj\}})$.
\begin{equation}
    \tr(\hat S_{\{ij\}}\hat S_{\{ik\}})=\tr(\hat S_i^2\hat S_{\{jk\}}+2\hat S_i\hat S_j\hat S_i\hat S_k).
\end{equation}
When $j=k\neq i$,
\begin{eqnarray}
    \tr(\hat S_{\{ij\}}^2)&=&\tr(2\hat S_i^2\hat S_j^2+2\hat S_i\hat S_j\hat S_i\hat S_j)\nonumber\\
    &=&\tr(4\hat S_i^2\hat S_j^2+2i\varepsilon_{ij\ell}\hat S_i\hat S_j\hat S_\ell).\nonumber
    \label{eq:A12}
\end{eqnarray}
The second term contributes
\begin{eqnarray}
    i\varepsilon_{ij\ell}\tr(2\hat S_i\hat S_j\hat S_\ell)&=&i\varepsilon_{ij\ell}\tr(\hat S_i\hat S_j\hat S_\ell+\hat S_i\hat S_j\hat S_\ell)\nonumber\\
    &=&i\varepsilon_{ij\ell}\tr(\hat S_i\hat S_j\hat S_\ell+\hat S_i\hat S_\ell\hat S_j\nonumber\\
    &&+i\varepsilon_{j\ell m}\hat S_i\hat S_m)\nonumber\\
    &=&-\varepsilon_{ij\ell}\varepsilon_{j\ell m}\tr(\hat S_i\hat S_m)\nonumber\\
    &=&-\tr(\hat S_i^2)=-2.
    \label{eq:A13}
\end{eqnarray}
To estimate the first term, we notice that
\begin{eqnarray}
    4\tr(\hat S_i^2\hat S_j^2)&=&-4i\varepsilon_{ij\ell}\tr(\hat S_i^2\hat S_j(\hat S_\ell\hat S_i-\hat S_i\hat S_\ell))\nonumber\\
    &=&-4i\varepsilon_{ij\ell}\tr(\hat S_i^3\hat S_j\hat S_\ell-\hat S_i^2\hat S_j\hat S_i\hat S_\ell)\nonumber\\
    &=&-4i\varepsilon_{ij\ell}\tr(\hat S_i^3\hat S_j\hat S_\ell-\hat S_i^3\hat S_j\hat S_\ell\nonumber\\
    &&+i\varepsilon_{ij\ell}\hat S_i^2\hat S_\ell^2)\nonumber\\
    &=&4\tr(\hat S_i^2\hat S_\ell^2),
\end{eqnarray}
which immediately gives $\tr(\hat S_1^2\hat S_2^2)=\tr(\hat S_2^2\hat S_3^2)=\tr(\hat S_3^2\hat S_1^2)$. Because
\begin{eqnarray}
\tr(\hat S_i^2\hat S_j^2)&=&\tr(\hat S_i^2(2\hat I_3-\hat S_i^2-\hat S_\ell^2))\nonumber\\
&=&2\tr(\hat S_i^2)-\tr(\hat S_i^4)-\tr(\hat S_i^2\hat S_\ell^2)\nonumber\\
&=&4-\tr(\hat S_i^4)-\tr(\hat S_i^2\hat S_j^2),
\end{eqnarray}
we have
\begin{equation}
    2\tr(\hat S_i^2\hat S_j^2)=4-\tr(\hat S_i^4).
    \label{eq:A15}
\end{equation}
For the same reason, $2\tr(\hat S_i^2\hat S_j^2)=4-\tr(\hat S_j^4)$. So we have $\tr(\hat S_1^4)=\tr(\hat S_2^4)=\tr(\hat S_3^4)$. Due to the $SO(3)$ rotation symmetry, $(\hat S_i+\hat S_j)/\sqrt2$ is also a normalized angular momentum operator, so we have
\begin{equation}
    \tr((\hat S_i+\hat S_j)^4/4)=\tr(\hat S_i^4)=\tr(\hat S_j^4),
\end{equation}
which immediately gives
\begin{eqnarray}
   && 2\tr(\hat S_i^4)+2\tr(\hat S_i^2\hat S_{\{ij\}})+2\tr(\hat S_j^2\hat S_{\{ij\}})\nonumber\\
    &&~~~~+2\tr(\hat S_i^2\hat S_j^2)+\tr(\hat S_{\{ij\}}^2)=4\tr(\hat S_i^4).
\end{eqnarray}
It is easy to see that
\begin{equation}
    \tr(\hat S_j^3\hat S_k)=-i\tr(\hat S_j^3\hat S_i\hat S_j-\hat S_j^3\hat S_j\hat S_i)=0.
\end{equation}
So $\tr(\hat S_i^2\hat S_{\{ij\}})=\tr(\hat S_j^2\hat S_{\{ij\}})=0$. And with Eq. (\ref{eq:A12}) and Eq. (\ref{eq:A13}), we have
\begin{equation}
    3\tr(\hat S_i^2\hat S_j^2)=1+\tr(\hat S_i^4).
\end{equation}
Together with Eq. (\ref{eq:A15}), we can get
\begin{equation}
    \tr(\hat S_i^4)=2,~~~~\tr(\hat S_i^2\hat S_j^2)=1.
\end{equation}
So when $j=k\neq i$
\begin{equation}
    \tr(\hat S_{\{ij\}}^2)=2
\end{equation}
When $j\neq k$, and $j,k\neq i$, we could assume that $(ijk)$ is an even permutation of $(1,2,3)$, then
\begin{eqnarray}
    \tr(\hat S_{\{ij\}}\hat S_{\{ik\}})&=&\tr(\hat S_i^2\hat S_{\{jk\}}+2\hat S_i\hat S_j\hat S_i\hat S_k)\nonumber\\
    &=&\tr(2\hat S_i^2\hat S_{\{jk\}})\nonumber\\
    &=&2\tr((2\hat I_3-\hat S_j^2-\hat S_k^2)\hat S_{\{jk\}})\nonumber\\
    &=&-2\tr((\hat S_j^2+\hat S_k^2)\hat S_{\{jk\}}).
\end{eqnarray}
So $\tr(\hat S_{\{ij\}}\hat S_{\{ik\}})=0$ when $j\neq k$, and $j,k\neq i$. When $i=k, i\neq j$,
\begin{equation}
    \tr(\hat S_{\{ii\}}\hat S_{\{ij\}})=4\tr(\hat S_i^3\hat S_j)=0.
\end{equation}
For $\tr(\hat S_{\{ii\}}\hat S_{\{jj\}})$, we have
\begin{equation}
    \tr(\hat S_{\{ii\}}\hat S_{\{jj\}})=4\tr(\hat S_i^2\hat S_j^2)=4.
\end{equation}
When $i=j$, $\tr(\hat S_{\{ii\}}^2)=8$, when $i\neq j$, $\tr(\hat S_{\{ii\}}\hat S_{\{jj\}})=4$.

As a summary, we have
\begin{align}
    \tr(\hat S_i)&=0,\\
    \tr(\hat S_i \hat S_j)&=2\delta_{ij},\\
    \tr(\hat S_i \hatSij{jk})&=0,\\
    \tr(\hat S_{\{ij\}}\hat S_{\{k\ell\}})&=2(\delta_{ik}\delta_{j\ell}+\delta_{i\ell}\delta_{jk})+4\delta_{ij}\delta_{k\ell}.
\end{align}
So the orthonormal condition requires
\begin{eqnarray}
    &&v_1^2+v_1v_2+v_2^2=1/8,\nonumber\\
    &&u_1^2+u_1u_2+u_2^2=1/8,\\
    &&2u_1v_1+2u_2v_2+u_1v_2+u_2v_1=0.\nonumber
\end{eqnarray}
The solution of these equations is just two orthogonal vectors whose norm is $1/2$ with the inner product defined by the quadratic form
\begin{equation}
    \left(
\begin{array}{cc}
 2 & 1 \\
 1 & 2 \\
\end{array}
\right).
\end{equation}
So we show that
\begin{eqnarray}
    &&\frac{1}{\sqrt3}\hat I_3,~\frac{1}{\sqrt2}\hat S_1,~\frac{1}{\sqrt2}\hat S_2,~\frac{1}{\sqrt2}\hat S_3,\nonumber\\
    &&\frac{1}{\sqrt2}\hat S_{\{12\}},~\frac{1}{\sqrt2}\hat S_{\{23\}},~\frac{1}{\sqrt2}\hat S_{\{31\}},\nonumber\\
    &&\frac{1}{\sqrt6}\Bigl[\hat S_{\{11\}}\cos\left(\alpha-\frac{\pi}{3}\right)-\hat S_{\{22\}}\sin\left(\alpha-\frac{\pi}{6}\right)\nonumber\\
    &&~~~~~~~~~ -\hat S_{\{33\}}\cos\alpha\Bigr],\nonumber\\
    &&\frac{1}{\sqrt6}\Bigl[-\hat S_{\{11\}}\sin\left(\alpha-\frac{\pi}{3}\right)-\hat S_{\{22\}}\cos\left(\alpha-\frac{\pi}{6}\right)\nonumber\\
    &&~~~~~~~~~ +\hat S_{\{33\}}\sin\alpha\Bigr],\nonumber\\
    &&\alpha\in[0,2\pi)
\end{eqnarray}
forms an orthonormal basis of $\mathfrak A_3$ under the inner product defined by $(\hat A,\hat B)\equiv\tr(\hat A^\dagger\hat B)$. In the basis of the eigenstates of $\hat S_3$, the matrix representation of this basis is
\begin{align} 
&\frac{1}{\sqrt3}\left(
\begin{array}{ccc}
 1 & 0 & 0 \\
 0 & 1 & 0 \\
 0 & 0 & 1 \\
\end{array}
\right),\nn 
&\frac{1}{2}\left(
\begin{array}{ccc}
 0 & 1 & 0 \\
 1 & 0 & 1 \\
 0 & 1 & 0 \\
\end{array}
\right),~\frac{1}{2}\left(
\begin{array}{ccc}
 0 & -i & 0 \\
 i & 0 & -i \\
 0 & i & 0 \\
\end{array}
\right),~\frac{1}{\sqrt2}\left(
\begin{array}{ccc}
 1 & 0 & 0 \\
 0 & 0 & 0 \\
 0 & 0 & -1 \\
\end{array}
\right),\nn
&\frac{1}{\sqrt2}\left(
\begin{array}{ccc}
 0 & 0 & -i \\
 0 & 0 & 0 \\
 i & 0 & 0 \\
\end{array} \right),~\frac{1}{2}\left(
\begin{array}{ccc}
 0 & -i & 0 \\
i & 0 & i \\
 0 & -i & 0 \\
\end{array} \right), ~\frac{1}{2}\left(
\begin{array}{ccc}
 0 & 1 & 0 \\
 1 & 0 & -1 \\
 0 & -1 & 0 \\
\end{array} \right), \nn
&\frac{1}{\sqrt6}\left(
\begin{array}{ccc}
 -\cos\alpha & 0 & \sqrt3\sin\alpha \\
 0 & 2\cos\alpha & 0 \\
 \sqrt3\sin\alpha & 0 & -\cos\alpha \\
\end{array} \right),\nn
&\frac{1}{\sqrt6}\left(
\begin{array}{ccc}
 \sin\alpha & 0 & \sqrt3\cos\alpha \\
 0 & -2\sin\alpha & 0 \\
 \sqrt3\cos\alpha & 0 & \sin\alpha \\
\end{array} \right).  
\end{align}
It is worth to emphasize that $\mathfrak A_3$ itself is not a real associative algebra under the matrix product, because the product of self-adjoint operators could not be self-adjoint operators.

It is also a common practice to expand $\rho_{W}$ with eight Gell-Mann matrices, as in Refs.~\cite{Barr:2021zcp,Aguilar-Saavedra:2022wam,Ashby-Pickering:2022umy}. The matrix representation of $\hat S_i$ and $\hatSij{ij}$ in the basis of the eigenstates of $\hat S_z$, and their relation with the Gell-Mann matrices $\lambda_a(a=1,\cdots,8)$ are
\begin{align}
&\hat S_x = \frac{1}{\sqrt{2}}(\lambda_1+\lambda_6) = \frac{1}{\sqrt{2}}\left(
\begin{array}{ccc}
 0 & 1 & 0 \\
 1 & 0 & 1 \\
 0 & 1 & 0 \\
\end{array}
\right), \nn
&\hat  S_y = \frac{1}{\sqrt{2}}(\lambda_2+\lambda_7)
= \frac{1}{\sqrt{2}}\left(
\begin{array}{ccc}
 0 & -i & 0 \\
 i & 0 & -i \\
 0 & i & 0 \\
\end{array}
\right), \nn
& \hat S_z = \frac{1}{2}\lambda_3 +  \frac{\sqrt{3}}{2}\lambda_8
= \left(
\begin{array}{ccc}
 1 & 0 & 0 \\
 0 & 0 & 0 \\
 0 & 0 & -1 \\
\end{array}
\right),\\
&\hat S_{\{xy\}} =  \lambda_5 
= \left(
\begin{array}{ccc}
 0 & 0 & -i \\
 0 & 0 & 0 \\
 i & 0 & 0 \\
\end{array} \right), \nn
& \hat S_{{\{xz\}}} = \frac{1}{\sqrt{2}}(\lambda_1 -\lambda_6)
=\frac{1}{\sqrt{2}}\left(
\begin{array}{ccc}
 0 & 1 & 0 \\
1 & 0 & -1 \\
 0 & -1 & 0 \\
\end{array} \right),  \nn
&\hat S_{\{yz\}}  = \frac{1}{\sqrt{2}}(\lambda_2 -\lambda_7)
=\frac{1}{\sqrt{2}}\left(
\begin{array}{ccc}
 0 & -i & 0 \\
 i & 0 & i \\
 0 & -i & 0 \\
\end{array} \right), \nn
&\hat S_{\{xx\}} = \frac{4}{3}I_3 - \frac{1}{2}\lambda_3 + \lambda_4 + \frac{1}{2\sqrt{3}}\lambda_8
=\left(
\begin{array}{ccc}
 1 & 0 & 1 \\
 0 & 2 & 0 \\
 1 & 0 & 1 \\
\end{array} \right),  \nn
&\hat S_{\{yy\}} = \frac{4}{3}I_3 - \frac{1}{2}\lambda_3 - \lambda_4 + \frac{1}{2\sqrt{3}}\lambda_8
=\left(
\begin{array}{ccc}
 1 & 0 & -1 \\
 0 & 2 & 0 \\
 -1 & 0 & 1 \\
\end{array} \right), \nn
&\hat S_{\{zz\}} = \frac{4}{3}I_3 + \lambda_3-\frac{1}{\sqrt{3}}\lambda_8
=\left(
\begin{array}{ccc}
 2 & 0 & 0 \\
 0 & 0 & 0 \\
 0 & 0 & 2 \\
\end{array}
\right).  
\end{align}

\section{Density Matrix Reconstruction}\label{app:densitymatrix}
\subsection{One $W$-boson}
Because the moving direction of the anti-fermion in the $W$-rest frame is just the measured spin direction of the $W$ boson, we have
\begin{equation}
    \hat \Pi_{\frakn}=\ket{S_\frakn=1}\bra{S_\frakn=1},
\end{equation}
which is not only a projective operator but also a density matrix of a pure state. So it could be represented by
\begin{equation}
    \hat \Pi_{\frakn}=\frac{1}{3}\hat I_3+\sum_{i=1}^3d(\frakn)_i\hat S_i+\sum_{i,j=1}^3q(\frakn)_{ij}\hat S_{\{ij\}}.
\end{equation}
Along any direction $\vec v$, the spin expectation value of this state is $2\vec v\cdot d(\frakn)$. Because it is the spin eigenstate along $\vec\frakn$ whose eigenvalue is 1, we have
\begin{eqnarray}
    d(\frakn)_i&=&\frac{\frakn_i}{2},\\
    \hat \Pi_{\frakn}&=&\frac{1}{3}\hat I_3+\frac{1}{2}\sum_{i=1}^3\frakn_i\hat S_i+\sum_{i,j=1}^3q(\frakn)_{ij}\hat S_{\{ij\}}.
\end{eqnarray}
For a density matrix of a pure state (a projective operator), by definition we have
\begin{equation}
    \hat\Pi_\frakn^2=\hat\Pi_\frakn.
\end{equation}
The square of the density matrix
\begin{eqnarray}
\hat\Pi_\frakn^2&=&\frac{1}{9}\hat I_3+\frac{1}{3}\sum_i^3\frakn_i\hat S_i+\sum_{i,j=1}^3\left[\frac{2}{3}q(\frakn)_{ij}+\frac{1}{8}\frakn_i\frakn_j\right]\hat S_{\{ij\}}\nonumber\\
&&+\frac{1}{2}\sum_{i,j,k=1}^3\frakn_iq(\frakn)_{jk}\left(\hat S_i\hat S_{\{jk\}}+\hat S_{\{jk\}}\hat S_i\right)\nonumber\\
&&+\frac{1}{2}\sum_{i,j,k,\ell=1}^3q(\frakn)_{ij}q(\frakn)_{k\ell}(\hat S_{\{ij\}}\hat S_{\{k\ell\}}+\hat S_{\{k\ell\}}\hat S_{\{ij\}}).\nonumber\\
&
\end{eqnarray}
With the normalization condition, we have $\tr(\hat\Pi_\frakn^2)=\tr(\hat\Pi_\frakn)=1$
\begin{eqnarray}
    1&=&\frac{1}{3}+4\sum_{i,j=1}^3\delta_{ij}\left[\frac{2}{3}q(\frakn)_{ij}+\frac{1}{8}\frakn_i\frakn_j\right]+\sum_{i,j,k,\ell=1}^3q(\frakn)_{ij}\nonumber\\
    &&\times q(\frakn)_{k\ell}[2(\delta_{ik}\delta_{j\ell}+\delta_{i\ell}\delta_{jk})+4\delta_{ij}\delta_{k\ell}]\nonumber\\
    &=&\frac{5}{6}+4\sum_{i,j=1}^3q(\frakn)_{ij}^2,\nonumber\\
    &\Rightarrow&\sum_{i,j=1}^3q(\frakn)_{ij}^2=\frac{1}{24}.
    \label{eq:B7}
\end{eqnarray}
Next we check the relations $\tr(\hat\Pi_\frakn^2\hat S_i)=\tr(\hat\Pi_\frakn\hat S_i)=\frakn_i$. Notice that the inner product $\tr(A^\dagger B)$ is invariant under the transformation $\hat I_3\to \hat I_3, \hat S_{\{ij\}}\to \hat S_{\{ij\}}, \hat S_i\to -\hat S_i$ ($i,j=1,2,3$), it is easy to see that $\tr((\hat S_{\{ij\}}\hat S_{\{k\ell\}}+\hat S_{\{k\ell\}}\hat S_{\{ij\}})\hat S_m)=0$, so
\begin{eqnarray}
    \frakn_i&=&\frac{2}{3}\sum_{j=1}^3\delta_{ij}\frakn_j+\frac{1}{2}\sum_{j,k,\ell=1}^3\frakn_jq(\frakn)_{k\ell}[2(\delta_{ik}\delta_{j\ell}+\delta_{i\ell}\delta_{jk})\nonumber\\
    &&+4\delta_{ij}\delta_{k\ell}]\nonumber\\
    &=&\frac{2\frakn_i}{3}+\sum_{j=1}^3[q(\frakn)_{ij}\frakn_j+q(\frakn)_{ji}\frakn_j+2q(\frakn)_{jj}\frakn_i],\nonumber\\
    &\Rightarrow&\sum_{j=1}^3q(\frakn)_{ij}\frakn_j=\frac{1}{6}\frakn_i.
    \label{eq:B8}
\end{eqnarray}
Finally, we check that $\tr(\hat\Pi_\frakn^2\hat S_{\{ij\}})=\tr(\hat\Pi_\frakn\hat S_{\{ij\}})=4q(\frakn)_{ij}$ for $i\neq j$.
\begin{eqnarray}
    4q(\frakn)_{ij}&=&\sum_{k,\ell=1}^3\left[\frac{2}{3}q(\frakn)_{k\ell}+\frac{1}{8}\frakn_k\frakn_\ell\right][2(\delta_{ik}\delta_{j\ell}+\delta_{i\ell}\delta_{jk})\nonumber\\
    &&+4\delta_{ij}\delta_{k\ell}]+\frac{1}{2}\sum_{a,b,c,d=1}^3q(\frakn)_{ab}q(\frakn)_{cd}\nonumber\\
    &&\times\tr[(\hat S_{\{ab\}}\hat S_{\{cd\}}+\hat S_{\{cd\}}\hat S_{\{ab\}})\hat S_{\{ij\}}].\nonumber\\
    &
\end{eqnarray}
To estimate the last trace, we notice that the inner product is invariant under the ``parity'' transformation: $\hat I_3\to \hat I_3$, $\hat S_{\{ij\}}\to (-1)^{N_a}\hat S_{\{ij\}}, \hat S_i\to (-1)^{N_a}\hat S_i$ for specific $a$, where $N_a$ is the times the number $a$ appears in the indices. So to have a non-vanished $\tr[(\hat S_{\{ab\}}\hat S_{\{cd\}}+\hat S_{\{cd\}}\hat S_{\{ab\}})\hat S_{\{ij\}}]$, each one of $1,2,3$ must appear even times in the indices. With this result and the symmetric structure, one could verify that
\begin{eqnarray}
&&\tr[(\hat S_{\{ab\}}\hat S_{\{cd\}}+\hat S_{\{cd\}}\hat S_{\{ab\}})\hat S_{\{ij\}}]\nonumber\\
&=&8\delta_{ij}(\delta_{ad}\delta_{bc}+\delta_{ac}\delta_{bd})+8(\delta_{ia}\delta_{jb}\delta_{cd}+\delta_{ib}\delta_{ja}\delta_{cd}\nonumber\\
&&+\delta_{ic}\delta_{jd}\delta_{ab}+\delta_{id}\delta_{jc}\delta_{ab})-2(\delta_{ia}\delta_{jc}\delta_{bd}+\delta_{ib}\delta_{jc}\delta_{ad}\nonumber\\
&&+\delta_{ia}\delta_{jd}\delta_{bc}+\delta_{ib}\delta_{jd}\delta_{ac}+
\delta_{ic}\delta_{ja}\delta_{bd}+\delta_{ic}\delta_{jb}\delta_{ad}\nonumber\\
&&+\delta_{id}\delta_{ja}\delta_{bc}+\delta_{id}\delta_{jb}\delta_{ac}).
\end{eqnarray}
So when $i\neq j$
\begin{equation}
    4q(\frakn)_{ij}=\frac{8}{3}q(\frakn)_{ij}+\frac{1}{2}\frakn_i\frakn_j-8\sum_{k=1}^3q(\frakn)_{ik}q(\frakn)_{kj},
    \label{eq:B11}
\end{equation}
and when $j=i$
\begin{eqnarray}
    \frac{4}{3}+4q(\frakn)_{ii}&=&\frac{4}{9}+\frac{8}{3}q(\frakn)_{ii}+\frac{1}{2}\frakn_i\frakn_i+\frac{1}{2}\nonumber\\
    &&+8\sum_{j,k=1}^3q(\frakn)_{jk}^2-8\sum_{j=1}^3q(\frakn)_{ij}^2,
\end{eqnarray}
The traceless solution of Eq.~(\ref{eq:B7}), Eq.~(\ref{eq:B8})  and Eq.~(\ref{eq:B11}) is
\begin{equation}
    q(\frakn)_{ij}=\frac{1}{4}\frakn_i\frakn_j-\frac{1}{12}\delta_{ij},
\end{equation}
So
\begin{eqnarray}
    \hat \Pi_{\frakn}&=&\frac{1}{3}\hat I_3+\frac{1}{2}\sum_{i=1}^3\frakn_i\hat S_i+\frac{1}{4}\sum_{i,j=1}^3\left(\frakn_i\frakn_j-\frac{1}{3}\delta_{ij}\right)\hat S_{\{ij\}}\nonumber\\
    &=&\frac{1}{2}(\hat S_\frakn+\hat S_\frakn^2),
\end{eqnarray}
where $\hat S_\frakn\equiv\sum_{i=1}^3\frakn_i\hat S_i$.

The probability of finding an anti-fermion in an infinitesimal solid angle $\dO$ of direction $\vec{\frakn}=(\sin\theta\cos\varphi,\sin\theta\sin\varphi,\cos\theta)$ from the $W$-boson decay products is
\begin{align}\label{eq:prhoW}
    p(\vec{\frakn};\rho_W)&=\frac{1}{N}\tr(\hat\rho_W \hat\Pi_{\frakn})\nn
    &=\frac{1}{N}\tr\Biggl[\left(\frac{1}{3}\hat I_3+d_i\hat S_i+q_{ij}\hat S_{\{ij\}}\right)\nn
    &~~~~\left(\frac{1}{3}\hat I_3+\frac{1}{2}\frakn_i\hat S_k+q(\frakn)_{k\ell}\hat S_{\{k\ell\}}\right)\Biggr]\nn
    &= \frac{1}{N} \left( \frac{1}{3}+d_i\frakn_i + q_{ij}q(\frakn)_{ij} \right)\nn
    &= \frac{1}{N} \Big( \frac{1}{3}+d_i\frakn_i + q_{ij} \frakn_i\frakn_j \Big)\nn.
    &=\frac{1}{N}\Big(\frac{1}{3} + d_1\sin\theta\cos\varphi  + d_2\sin\theta\sin\varphi  \nn
    &~~~~+ d_3\cos\theta + q_{12}\sin^2\theta\sin2\varphi + q_{31}\sin2\theta\cos\varphi \nn
    &~~~~+ q_{23}\sin2\theta\sin\varphi + q_{11}\sin^2\theta\cos^2\varphi \nn 
    &~~~~+ q_{22}\sin^2\theta\sin^2\varphi + q_{33}\cos^2\theta\Big).
\end{align}
The normalization constant $N$ is calculated with 
\begin{equation}
    \int p(\vec{\frakn};\rho_W) \dO =1,
\end{equation}
which gives $N=4\pi/3$.
The averages of these kinetic observables $\frakn_i$ (the $i$th component of the normalized 3-vector of the moving direction of the anti-fermion in the rest frame of the $W$-boson, or equivalently the $i$th component of the spin of the $W$-boson) and $\frakq_{ij}$ (the correlations between the spin components) give the information of the density matrix 
\begin{align}
    \label{eq:averagesd1}
    \av{\frakn_1}&= \int \sin\theta \cos\varphi ~p(\vec{\frakn};\rho_W) \dO \nn
    &= \frac{3}{4\pi} \int^{\pi}_{0} d_1 \sin^3\theta \dtheta \int^{2\pi}_{0} \cos^2\varphi \dphi \nn
    &= d_1, \\
    \av{\frakn_2}&= \int \sin\theta \sin\varphi ~p(\vec{\frakn};\rho_W) \dO \nn
    &= \frac{3}{4\pi} \int^{\pi}_{0} d_2 \sin^3\theta \dtheta \int^{2\pi}_{0} \sin^2\varphi \dphi \nn
    &= d_2, \\
    \av{\frakn_3}&= \int \cos\theta ~p(\vec{\frakn};\rho_W) \dO \nn
    &= \frac{3}{4\pi} \int^{\pi}_{0} d_3 \cos^2\theta\sin\theta \dtheta \int^{2\pi}_{0} \dphi \nn
    &= d_3, \\
    \av{\frakq_{11}}&\equiv \int \left(\sin^2\theta \cos^2\varphi - \frac{1}{3}\right) ~p(\vec{\frakn};\rho_W) \dO \nn
    &= \frac{1}{4\pi} \iint \left(\sin^2\theta \cos^2\varphi - \frac{1}{3}\right) (1 + 3q_{11}\sin^2\theta\cos^2\varphi \nn
    &+ 3q_{22}\sin^2\theta\sin^2\varphi +3q_{33}\cos^2\theta) \sin\theta \dtheta\dphi \nn
    &= \frac{2}{5} q_{11}, \\
    \av{\frakq_{22}}&\equiv \int \left(\sin^2\theta \sin^2\varphi - \frac{1}{3}\right) ~p(\vec{\frakn};\rho_W) \dO \nn
    &= \frac{1}{4\pi} \iint \left(\sin^2\theta \sin^2\varphi - \frac{1}{3}\right) (1 + 3q_{11}\sin^2\theta\cos^2\varphi \nn
    &+ 3q_{22}\sin^2\theta\sin^2\varphi +3q_{33}\cos^2\theta) \sin\theta \dtheta\dphi \nn
    &= \frac{2}{5} q_{22}, \\
    \av{\frakq_{33}}&\equiv \int \left(\cos^2\theta - \frac{1}{3}\right) ~p(\vec{\frakn};\rho_W) \dO \nn
    &= \frac{1}{4\pi} \iint \left(\cos^2\theta - \frac{1}{3}\right) (1 + 3q_{11}\sin^2\theta\cos^2\varphi \nn
    &+ 3q_{22}\sin^2\theta\sin^2\varphi +3q_{33}\cos^2\theta) \sin\theta \dtheta\dphi \nn
    &= \frac{2}{5} q_{33}, \\
    \av{\frakq_{12}}&= \int \sin^2\theta \sin\varphi \cos\varphi ~p(\vec{\frakn};\rho_W) \dO \nn
    &= \frac{3}{4\pi} \int^{\pi}_{0} 2q_{12} \sin^5\theta \dtheta \int^{2\pi}_{0} \sin^2\varphi \cos^2\varphi \dphi \nn
    &= \frac{2}{5} q_{12}, \\
    \av{\frakq_{23}}&= \int \sin\theta \cos\theta \sin\varphi ~p(\vec{\frakn};\rho_W) \dO \nn
    &= \frac{3}{4\pi} \int^{\pi}_{0} 2q_{23} \sin^3\theta \cos^2\theta \dtheta \int^{2\pi}_{0} \sin^2\varphi \dphi \nn
    &= \frac{2}{5} q_{23}, \\
    \label{eq:averagesq13}
    \av{\frakq_{13}}&= \int \sin\theta \cos\theta \cos\varphi ~p(\vec{\frakn};\rho_W) \dO \nn
    &= \frac{3}{4\pi} \int^{\pi}_{0} 2q_{13} \sin^3\theta \cos^2\theta \dtheta \int^{2\pi}_{0} \cos^2\varphi \dphi \nn
    &= \frac{2}{5} q_{13}.
\end{align}
Eqs.~\eqref{eq:averagesd1}-\eqref{eq:averagesq13} can be summarized as
\begin{align}
    d_i = \av{\frakn_i},\quad
    q_{ij} = \frac{5}{2}\av{\frakq_{ij}}.
\end{align}
It is found that the parameters $d_i$ and $q_{ij}$, which are related to the circular polarization (spin eigenstates) and linear polarization of the $W$ boson, are determined by the dipole and quadrupole distributions of the anti-fermion respectively. To see this fact clearly and quickly, we notice that the basis of the matrix representation of the $SO(3)$ group we used is the eigenstates of the rotation transformation around the 3rd axis. However, in the vector representation (under the basis of the linear polarization states $\ket{j},~j=1,2,3$),
\begin{eqnarray}
    &&\hat S_1=\left(\begin{matrix}
    0 & 0 & 0\\
    0 & 0 & -i\\
    0 & i & 0
    \end{matrix}\right),
    \hat S_2=\left(\begin{matrix}
    0 & 0 & i\\
    0 & 0 & 0\\
    -i & 0 & 0
    \end{matrix}\right),
    \hat S_3=\left(\begin{matrix}
    0 & -i & 0\\
    i & 0 & 0\\
    0 & 0 & 0
    \end{matrix}\right),\nn
    &&\hat S_{\{12\}}=\left(\begin{matrix}
    0 & -1 & 0\\
    -1 & 0 & 0\\
    0 & 0 & 0
    \end{matrix}\right),~~~~\hat S_{\{23\}}=\left(\begin{matrix}
    0 & 0 & 0\\
    0 & 0 & -1\\
    0 & -1 & 0
    \end{matrix}\right),\nn
    &&\hat S_{\{31\}}=\left(\begin{matrix}
    0 & 0 & -1\\
    0 & 0 & 0\\
    -1 & 0 & 0
    \end{matrix}\right),~~~~\hat S_{\{11\}}=\left(\begin{matrix}
    0 & 0 & 0\\
    0 & 2 & 0\\
    0 & 0 & 2
    \end{matrix}\right),\nn
    &&\hat S_{\{22\}}=\left(\begin{matrix}
    2 & 0 & 0\\
    0 & 0 & 0\\
    0 & 0 & 2
    \end{matrix}\right),~~~~\hat S_{\{33\}}=\left(\begin{matrix}
    2 & 0 & 0\\
    0 & 2 & 0\\
    0 & 0 & 0
    \end{matrix}\right).
\end{eqnarray}
So (only keep the relative phase between the normalized state vectors) the eigenvectors of $\hat S_3$ are
\begin{eqnarray}
|\uparrow\rangle&=&-\frac{1}{\sqrt2}(|1\rangle+i|2\rangle),\nn
|0\rangle&=&|3\rangle,\nn
|\downarrow\rangle&=&\frac{e^{-i\delta}}{\sqrt2}(|1\rangle-i|2\rangle),\nonumber
\end{eqnarray}
so
\begin{eqnarray}
|1\rangle&=&\frac{1}{\sqrt2}(-|\uparrow\rangle+e^{i\delta}|\downarrow\rangle),\nn
|2\rangle&=&\frac{i}{\sqrt2}(|\uparrow\rangle+e^{i\delta}|\downarrow\rangle),\nonumber
\end{eqnarray}
which immediately gives
\begin{eqnarray}
\hat\Pi_{11}&=&\frac{1}{3}\hat I_3+\frac{1}{6}(-2\hat S_{\{11\}}+\hat S_{\{22\}}+\hat S_{\{33\}})\nn
&&+\frac{1}{2}\hat S_{\{11\}}-\left(\hat S_1\cos\frac{\delta}{2}+\hat S_2\sin\frac{\delta}{2}\right)^2\nn
&=&\hat I_3-\left(\hat S_1\cos\frac{\delta}{2}+\hat S_2\sin\frac{\delta}{2}\right)^2,\nn
\hat\Pi_{22}&=&\frac{1}{3}\hat I_3+\frac{1}{6}(\hat S_{\{11\}}-2\hat S_{\{22\}}+\hat S_{\{33\}})\nn
&&+\frac{1}{2}\hat S_{\{22\}}-\left(\hat S_2\cos\frac{\delta}{2}-\hat S_1\sin\frac{\delta}{2}\right)^2\nn
&=&\hat I_3-\left(\hat S_2\cos\frac{\delta}{2}-\hat S_1\sin\frac{\delta}{2}\right)^2,\nn
\hat\Pi_{33}&=&\frac{1}{3}\hat I_3+\frac{1}{6}(\hat S_{\{11\}}+\hat S_{\{22\}}-2\hat S_{\{33\}})\nn
&=&\hat I_3-\frac{1}{2}\hat S_{\{33\}}.
\end{eqnarray}
It is easy to see that the non-zero phase factor $\delta$ reflects the difference between the phase of the left-hand and right-hand circular polarization eigenstates, which could be removed by a rotation around the 3rd axis. Generally, the density matrix operator of the linear polarization state along the direction $\vec \frakn$ is
\begin{equation}
    \hat\Pi_{\frakn\frakn}=\hat I_3-\hat S_{\frakn}^2=\hat\Pi_{(-\frakn)(-\frakn)},
\end{equation}
which does not depend on the sign of $\frakn$. It is easy to check that the components of the direction of the linear polarization could be written as
\begin{equation}
    \frakn_i^2=1-\frac{1}{2}\tr(\hat \Pi_{\frakn\frakn}\hat S_{\{ii\}}).
\end{equation}

\subsection{$W$-boson pair}
Likewise, the density matrix of $W^\pm$ pair system can be reconstructed from the distribution of their decay products. In their rest frame of $W^\pm$ respectively, we use $\vec{\frakn}^\pm$ to denote the normalized directions of two outgoing anti-fermions decayed from $W^\pm$. The probability of finding a pair of anti-fermions along directions $\vec\frakn^\pm$ is
\begin{align}\label{eq:pnnrhoww}
    p(\vec{\frakn}^+,\vec{\frakn}^-&;\rho_{WW})\nn
    =\frac{1}{N^2} &\tr\left[\hat\rho_{WW} \Proj{S_{\frakn^+}=1}\otimes\Proj{S_{\frakn^-}=1}\right]\nn
    =\frac{1}{N^2} &\Big[ \frac{1}{9}+\frac{1}{3}(d_i^+ \frakn_i^+ + d_i^-\frakn_i^-)\nn 
    &+ \frac{1}{3}( q_{ij}^+ \frakq_{ij}^+ + q_{ij}^- \frakq_{ij}^-)\nn
    &+ C^d_{ij}\frakn_i^+\frakn_j^- + C^q_{ij,kl}\frakq_{ij}^+\frakq_{kl}^- \nn
    &+ C^{dq}_{i,jk}\frakn_i^+\frakq_{jk}^-  + C^{qd}_{ij,k}\frakq_{ij}^+\frakn_k^-  \Big],
\end{align}
where $N=4\pi/3$ is normalization constant as in Eq.~\eqref{eq:prhoW}.
As the density matrix of $W^\pm$ pair system is the direct product of the two subsystems, the parameters can be obtained by calculating the averages of the kinetic observables $\frakn_i^\pm$, $\frakq_{ij}^\pm$ and their combinations similarly. The average of an observable $X$ is calculated by
\begin{align}
    \av{X}&= \int X ~p(\vec{\frakn}^+,\vec{\frakn}^-;\rho_{WW}) \dO^+ \dO^-,
\end{align}
where $\dO^\pm$ denotes infinitesimal solid angles to find the anti-fermion direction $\vec{\frakn}^\pm(\theta^\pm,\phi^\pm)$ from the $W^\pm$-boson decay products in the rest frame of $W^\pm$, respectively.

Note that Eq.~\eqref{eq:pnnrhoww} can be factorized as
\begin{align}
    &p(\vec{\frakn}^+,\vec{\frakn}^-;\rho_{WW})\nn
    =&\frac{1}{N^2}\Big[ \frac{1}{3}\left(\frac{1}{3}+d_i^-\frakn_i^-+q_{ij}^-\frakq_{ij}^-\right) \nn
    &\quad+ \left(\frac{d_i^+}{3}+C_{ij}^d \frakn_j^- + C_{i,jk}^{dq}\frakq_{jk}^-\right)\frakn_i^+ \nn
    &\quad+ \left(\frac{q_{ij}^+}{3}+C_{ij,k}^{qd} \frakn_k^- + C_{ij,kl}^{q}\frakq_{jk}^-\right)\frakq_{ij}^+
    \Big]
\end{align}
Performing the integrals in Eqs.~\eqref{eq:averagesd1}-\eqref{eq:averagesq13} twice, we obtain
\begin{align}
    \av{\frakn^\pm_i}&=d^\pm_{i}, \\
    \av{\frakq^\pm_{ij}}&=\frac{2}{5}q^\pm_{ij},\\
    \av{\frakn^+_i \frakn^-_j}&=C^d_{ij}, \\
    \av{\frakq^+_{ij} \frakq^-_{kl}}&=\frac{4}{25}C^q_{ij,kl},\\
    \av{\frakn^+_{i}\frakq^-_{jk}}&=\frac{2}{5}C^{dq}_{i,jk},\\
    \av{\frakq^+_{ij}\frakn^-_{k}}&=\frac{2}{5}C^{qd}_{ij,k}.
\end{align}

\section{Calculation of the Bell observable}\label{app:observable}
The observable $\mathcal{I}_3$ in Eq.~\eqref{eq:ISL} is constructed from the probabilities derived from the measurement of the operators,
\begin{align}
    & \mathcal{I}_3(\hat{S}_{\vec{a}_1},\hat{S}_{\vec{a}_2};\hatSij{x_3 y_3},\hatSij{x_4 y_4}) \nn
    & \equiv + \big[P(S_{\vec{a}_1}=S_{\{x_3 y_3\}})+P(S_{\{x_3 y_3\}}=S_{\vec{a}_2}+1) \nn
    &\quad+P(S_{\vec{a}_2}=S_{\{x_4 y_4\}})+P(S_{\{x_4 y_4\}}=S_{\vec{a}_1})\big] \nn
    &\quad -\big[P(S_{\vec{a}_1}=S_{\{x_3 y_3\}}-1)+P(S_{\{x_3 y_3\}}=S_{\vec{a}_2}) \nn
    & \quad+P(S_{\vec{a}_2}=S_{\{x_4 y_4\}}-1)+P(S_{\{x_4 y_4\}}=S_{\vec{a}_1}-1)\big].
\end{align}
A direct way to evaluate $\mathcal{I}_3(\hat{S}_{\vec{a}_1},\hat{S}_{\vec{a}_2};\hatSij{x_3 y_3},\hatSij{x_4 y_4})$ is to project the density matrix $\hat\rho_{WW}$ to the eigenstates of the operators $\hat{S}_{\vec{a}_1}$, $\hat{S}_{\vec{a}_2}$, $\hatSij{x_3 y_3}$ and $\hatSij{x_4 y_4}$. 
For example, the first term of $\mathcal{I}_3(\hat{S}_{\vec{a}_1},\hat{S}_{\vec{a}_2};\hatSij{x_3 y_3},\hatSij{x_4 y_4})$ is
\begin{align}
    P(S_{\vec{a}_1}=S_{\{x_3 y_3\}})= \sum_{\lambda=-1}^1 & \tr\left[ \hat\rho_{WW} \Proj{S_{\vec{a}_1}=\lambda,S_{\{x_3 y_3\}}=\lambda } \right].
\end{align}
The projective operators of the spin operator $\hat{S}_{\vec{a}_1} (\hat S_{\vec{a}_1}\equiv\sum_{i=1}^3\vec{a}_{1i}\hat S_i)$ are
\begin{align}
    & \hat \Pi_{\vec{a}_1}(S_{\vec{a}_1}=-1)=\frac{1}{2}(-\hat S_{\vec{a}_1}+\hat S_{\vec{a}_1}^2), \\
    & \hat \Pi_{\vec{a}_1}(S_{\vec{a}_1}=1)=\frac{1}{2}(\hat S_{\vec{a}_1}+\hat S_{\vec{a}_1}^2), \\
    & \hat \Pi_{\vec{a}_1}(S_{\vec{a}_1}=0)=\hat I_3-\hat S_{\vec{a}_1}^2.
\end{align}
And the projective operators of the $\hatSij{x_3 y_3}$  are (see Eq.~\eqref{eq:eigSxy})
\begin{align}
    & \hat\Pi_{x_3 y_3}(S_{\{x_3 y_3\}}=-1)=\hat I_3-\hat S_{\vec{\epsilon}_{1}}^2, \; \vec{\epsilon}_{1}=\frac{\hat{x}_3+\hat{y}_3}{\sqrt{2}},\\
    & \hat\Pi_{x_3 y_3}(S_{\{x_3 y_3\}}=1)=\hat I_3-\hat S_{\vec{\epsilon}_{2}}^2, \quad \vec{\epsilon}_{2}=\frac{\hat{x}_3-\hat{y}_3}{\sqrt{2}},\\
    & \hat\Pi_{x_3 y_3}(S_{\{x_3 y_3\}}=0)=\hat I_3-\hat S_{\vec{\epsilon}_{3}}^2, \quad
    \vec{\epsilon}_3 = \hat{x}_3 \times \hat{y}_3.
\end{align}
Here, the $\hat{x}_3$ and $\hat{y}_3$ are the normalized directions of the two axes in the orthogonal coordinates $(\hat{x}_3,\hat{y}_3,\hat{z}_3)$.
Therefore the probability can be evaluated by
\begin{align}
    & P(S_{\vec{a}_1}=S_{\{x_3 y_3\}})  = \sum_{\lambda=-1}^1 \tr\left[ \hat\rho_{WW} \Proj{S_{\vec{a}_1}=\lambda,S_{\{x_3 y_3\}}=\lambda } \right] \nn
    & = \tr\left[\hat\rho_{WW} \cdot \hat \Pi_{\vec{a}_1}(S_{\vec{a}_1}=-1) \otimes \hat\Pi_{x_3 y_3}(S_{\{x_3 y_3\}}=-1) \right] \nn
    & \ +\tr\left[\hat\rho_{WW} \cdot \hat \Pi_{\vec{a}_1}(S_{\vec{a}_1}=1) \otimes \hat\Pi_{x_3 y_3}(S_{\{x_3 y_3\}}=1) \right] \nn
    & \ +\tr\left[\hat\rho_{WW} \cdot \hat \Pi_{\vec{a}_1}(S_{\vec{a}_1}=0) \otimes \hat\Pi_{x_3 y_3}(S_{\{x_3 y_3\}}=0) \right] \nn
    & = \ 1 - 2q_{ij}^- \epsilon_{3i} \epsilon_{3j} - 2 C_{i,jk}^{dq} a_{1i} (\epsilon_{1j} \epsilon_{1k} - \epsilon_{2j} \epsilon_{2k}) \nn
    & \quad +2 C^q_{ij,kl} a_{1i}a_{1j} (-\epsilon_{1k} \epsilon_{1l}-\epsilon_{2k} \epsilon_{2l}+ 2\epsilon_{3k} \epsilon_{3l}).
\end{align}
Similar to the definition of $\vec{\epsilon}_{1,2,3}$, we use $\vec{\omega}_{1}$, $\vec{\omega}_{2}$ and $\vec{\omega}_{3}$ to denote the polarization directions of the eigenstates of $\hatSij{x_4 y_4}$, and then evaluate the other terms. 
The explicit form of $\mathcal{I}_3$ is
\begin{align}
    & \mathcal{I}_3(\hat{S}_{\vec{a}_1},\hat{S}_{\vec{a}_2};\hatSij{x_3 y_3},\hatSij{x_4 y_4}) \nn
    & = 2 q^-_{ij}(\omega_{1i}\omega_{1j}+\omega_{2i}\omega_{2j}-2\omega_{3i}\omega_{3j}) \nn
    &\ + 2C^{dq}_{i,jk} a_{1i} (2\epsilon_{1j}\epsilon_{1k} - \epsilon_{2j}\epsilon_{2k} - \epsilon_{3j}\epsilon_{3k} + \omega_{1j}\omega_{1k} \nn
    & \qquad -2\omega_{2j}\omega_{2k} +\omega_{3j}\omega_{3k})\nn
    &\ + 2C^{dq}_{i,jk} a_{2i} (-2\epsilon_{1j}\epsilon_{1k} + \epsilon_{2j}\epsilon_{2k} + \epsilon_{3j}\epsilon_{3k} + 2\omega_{1j}\omega_{1k} \nn
    & \qquad- \omega_{2j}\omega_{2k} - \omega_{3j}\omega_{3k})\nn
    &\ + 6C^q_{ij,kl}a_{1i}a_{1j} (-\epsilon_{2k}\epsilon_{2l} + \epsilon_{3k}\epsilon_{3l} - \omega_{1k}\omega_{1l} + \omega_{3k}\omega_{3l}) \nn
    &\ + 6C^q_{ij,kl}a_{2i}a_{2j} (\epsilon_{2k}\epsilon_{2l} - \epsilon_{3k}\epsilon_{3l} - \omega_{2k}\omega_{2l} + \omega_{3k}\omega_{3l})
\end{align}
Given that the directions $\vec{a}_1$, $\vec{a}_2$, $\{x_3 y_3\}$ and $\{x_4 y_4\}$ of these operators chosen for the observable are not predetermined, it is necessary to perform a parameter scan over all possible directions. 
This enables us to identify the maximum of $\mathcal{I}_3$ as the observable.

\bibliographystyle{apsrev}
\bibliography{ref}

\end{document}